\documentclass[ preprint, showpacs,preprintnumbers, amsmath,amssymb, aps, prb]{revtex4-1}

\usepackage{graphicx}
\usepackage{dcolumn}
\usepackage{bm}
\usepackage{hyperref}
\usepackage[cp1250]{inputenc}
\usepackage[usenames,dvipsnames]{color}

\newcommand{\tr}{\textcolor{Red}}
\renewcommand{\tr}{}

\newcommand{\bk}{\mathbf{k}}

\newcommand{\bQ}{\mathbf{Q}}

\newcommand{\ua}{\uparrow}
\newcommand{\da}{\downarrow}
\newcommand{\ra}{\rightarrow}
\newcommand{\mb}{\mathbf}
\newcommand{\dg}{^\dagger}

\newcommand{\eq}{\begin{equation}}
\newcommand{\eqx}{\end{equation}}
\newcommand{\eqn}{\begin{eqnarray}}
\newcommand{\eqnx}{\end{eqnarray}}
\newcommand{\half}{\frac{1}{2}}
\newcommand{\ran}{\rangle}
\newcommand{\lan}{\langle}
\newcommand{\s}{\sigma}
\newcommand{\lm}{\lambda}
\newcommand{\D}{\Delta}

\begin{document}

\preprint{APS/123-QED}

\title{Coexistence of antiferromagnetism and superconductivity within $t$-$J$ model with strong correlations and nonzero spin polarization}

\author{Jan Kaczmarczyk}
\email{jan.kaczmarczyk@uj.edu.pl}
\affiliation{
Marian Smoluchowski Institute of Physics, Jagiellonian University, \linebreak Reymonta 4, 30-059 Kraków, Poland
}

\author{Jozef Spałek}
 \email{ufspalek@if.uj.edu.pl}
\affiliation{
Marian Smoluchowski Institute of Physics, Jagiellonian University, \linebreak Reymonta 4, 30-059 Kraków, Poland \\
Faculty of Physics and Applied Computer Science, AGH University of Science and Technology, Reymonta 19, 30-059 Kraków, Poland
}

%

\date{\today}

\begin{abstract}
The coexistence of antiferromagnetism with superconductivity is studied theoretically within the $t$-$J$ model with the Zeeman term included. The strong electron correlations are accounted for by means of the extended Gutzwiller projection method within a statistically-consistent approach proposed recently. The phase diagram on the band filling - magnetic field plane is shown, and subsequently the system properties are analyzed for the fixed band filling $n=0.97$. In this regime, the results reflect principal qualitative features observed recently in selected heavy fermion systems. Namely, (i) with the increasing magnetic field the system evolves from coexisting antiferromagnetic-superconducting phase, through antiferromagnetic phase, towards \tr{polarized paramagnetic state}, and (ii) the onset of superconducting order suppresses partly the staggered moment. The superconducting gap has both the spin-singlet and the staggered-triplet components, a direct consequence of a coexistence of the superconducting state with antiferromagnetism.
\end{abstract}


\maketitle

\section{Introduction}

The interplay of antiferromegnetism (AF) with superconductivity (SC) is one of the important topics in condensed-matter physics, \cite{Rev_Demler} as better understanding of this subject would improve our knowledge of a number of systems such as high-Tc, \cite{Rev_High_Tc} heavy-fermion, \cite{Pfleiderer} and organic \cite{Rev_Organic_SC} superconductors. In all those systems, superconductivity appears in the vicinity of magnetic phases (mostly antiferromagnetic, but also ferromagnetic \cite{Saxena, Aoki}). Moreover, magnetic interactions or fluctuations are very frequently considered to be the pairing mechanism in unconventional superconductors. \cite{Monthoux, MonthouxSF, MonthouxSF2} Typically, antiferromagnetism and superconductivity are competing quantum phenomena because of the competition between the Meissner-supercurrent screening and the internal-field generation by magnetic ordering. This antagonism can be overcome by a spatial separation of the AF and the SC phases or by subdivision of the $f$ electrons into more localized (resulting in AF) and more itinerant (participating in SC) parts. However, especially interesting is the situation, when the same electrons are involved in both phenomena, as is the case for some heavy-fermion systems. There, SC and AF can coexist easily, when the periodicity of magnetic structure $\lambda_{AF} (=2a)$ is much smaller than the coherence length $\xi$ for the Cooper pair. In other words, when $\xi \gg a$, the staggered exchange field averages out to zero within the coherence volume. In this respect the Ce-based '115' heavy-fermion compounds - the family of CeMIn$_5$ (with $M = Co, Rh, Ir$) \cite{Sarrao, Thompson, Shishido} is the most promising, as both antiferromagnetism and superconductivity are believed to arise from $4f$ electrons, where even the interplay of the two orders can be studied by tuning the system with pressure, magnetic field, or doping.

Also, recently a competitive coexistence of AF and SC has been reported in both CeRhIn$_5$ \cite{Chen, Knebel, Knebel2, ParkPNAS} and CeCo(In$_{1-x}$Cd$_{x}$)$_5$. \cite{Nair, Urbano} In the latter system, a mutual influence of AF and SC has been observed. Namely, it turns out that the onset of SC order with lowering temperature prevents any further increase of the antiferromagnetic magnetization. \cite{Nair} A similar type of coexistence has also been observed in CeRhSi$_3$. \cite{Kimura}

Generally, in the heavy-fermion systems strong correlations among electrons are the reason for an emergence of new and nontrivial physics. Those nontrivial features should be properly accounted for when modeling those systems. In the present paper, an investigation of the coexistence of AF with SC in an applied magnetic field is presented. To account for strong electron correlations, the Gutzwiller-projected $t$-$J$ model is used with the Zeeman term included. The extended Gutzwiller scheme proposed recently \cite{Fukushima} is utilized for calculation of statistical averages of the relevant operators. \tr{Our model, although at first sight seems too simplified to be related to heavy fermion systems, it nonetheless reflects qualitatively principal features observed recently in selected heavy fermion systems.}

It is commonly believed that the minimal model for investigation of heavy-fermion systems should be the two-band Periodic Anderson Model (PAM)  (see e.g. Ref.~\onlinecite{Sacramento}) or the Kondo lattice model. \cite{Tsunetsugu} On the other hand, the one-band calculations have already proved fruitful in the analysis of AF and SC coexistence in CeRhIn$_5$, \cite{Alvarez} as well as in investigations of the high-field low-temperature unconventional superconducting phase of CeCoIn$_5$. \cite{Yanase, Aperis} The narrow-band limit of PAM has been discussed  theoretically also elsewhere (see Refs.~\onlinecite{JS_RSP1} and \onlinecite{Maska}, Appendix A). Generally, it appears when only a single hybridized band is involved and in the heavy-fermion limit (i.e. when $f$-level occupancy $n_f = 1-\delta$, with $\delta \ll 1$). \tr{Simply put, the $t$-$J$-type model reflects the physics of those hybridized and strongly correlated systems in the narrow $f$-band limit.}

The paper is organized as follows. In Sec.~\ref{sec:model} we present the general theoretical formulation. In Sec.~\ref{sec:results} we show the numerical results, and finally in Sec.~\ref{sec:summary}, our findings are summarized and an outlook is provided.

\section{Model, order parameters, and constraints} \label{sec:model}

\begin{figure}
\begin{center}
\includegraphics[height=3cm]{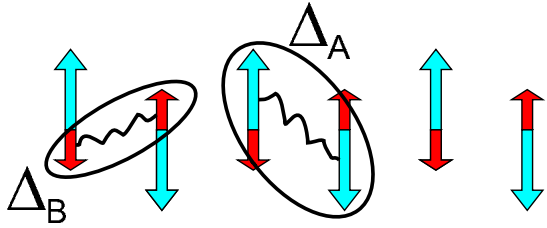}
\end{center}
\caption{(Color online) Spin-majority (blue, bigger arrows) and spin-minority (red, smaller arrows) electron spins in a system with the AF order and real-space superconducting gaps. $\Delta_{A}$ binds two spin-majority electrons, and $\Delta_{B}$ binds two spin-minority electrons and therefore, there is a priori no reason for these two gaps to coincide (as would be the case for no staggered $\pi$-triplet component). In other words, the two distinct gaps make effectively the $\ua-\da$ and $\da-\ua$ pairing components of the opposite-spin pairs distinguishable.}
\label{fig1}
\end{figure}

We start from the $t$-$J$ model with the Zeeman term included, as represented by the Hamiltonian\footnote{For the discussion of various forms of $t$-$J$ model see Ref. \onlinecite{JJ2} below.}
\eq
\hat{\mathcal{H}}_{tJ} = \hat{P} \Big( \sum_{ij\sigma} t_{ij} c_{i\sigma}\dg c_{j\sigma} + J \sum_{\langle ij \rangle} \mathbf{S_i S_j} - h \sum_{i\sigma} \sigma \hat{n}_{i \sigma} \Big) \hat{P}, \label{eq:tJ}
\eqx
where ${\langle ij \rangle}$ denotes the summation over bonds, and $\sigma = \pm 1$ is the spin $z$-component. \tr{Since its derivation, \cite{ChaoJS, ZR} the $t$-$J$ model represents an active field of research (see e.g. Ref. \onlinecite{Moreno} for a recent analysis of the one-dimensional situation). The $t$-$J$ model captures the essential ingredients of physics of the high-Tc superconductors.} The advantage of using this model is that both AF and SC come from a microscopic parameter - antiferromagnetic exchange $J$ and therefore there are no phenomenological terms in the Hamiltonian (as opposed to some earlier studies of AF and SC coexistence). We neglect the orbital effects, as the Maki parameter \cite{Maki} in the systems of our interest here is high. \cite{Kumagai2, Knebel2} The Gutzwiller projector $\hat{P} \equiv \Pi_i (1 - \hat{n}_{i\ua} \hat{n}_{i\da})$ eliminates double occupancies in real space. In the following we will use the more general correlator
\eq
\hat{P}_C \equiv \Pi_i \lambda_{i\ua}^{\hat{n}_{i\ua}/2} \lambda_{i\da}^{\hat{n}_{i\da}/2} (1 - \hat{n}_{i\ua} \hat{n}_{i\da}),
\label{eq:correlator}
\eqx
where $\lambda_{i\s}$ are the so-called fugacity factors. Also, this correlator connects the correlated $|\Psi \ran$ and uncorrelated $|\Psi_0 \ran$ wave functions, \cite{Anderson} via
\eq
| \Psi \ran = \hat{P}_C | \Psi_0 \ran.
\eqx
This allows to express average of any operator $\hat{O}$ in the correlated state as
\eq
\lan \hat{O} \ran \equiv \lan \Psi | \hat{O} | \Psi \ran = \frac{\lan \hat{P}_C \hat{O} \hat{P}_C \ran_0}{\lan \hat{P}_C \hat{P}_C \ran_0}, \label{eq:ave}
\eqx
where $\lan ... \ran_0 \equiv \lan \Psi_0| ... |\Psi_0 \ran$. With the above equation one can in principle calculate average value of Hamiltonian (\ref{eq:tJ}), namely
\eq
W \equiv \langle \hat{\mathcal{H}}_{tJ} \rangle = \sum_{ij\sigma} t_{ij} \langle c_{i\sigma}\dg c_{j\sigma} \rangle + J \sum_{\langle ij \rangle} ({\langle S_i^z S_j^z \rangle} + {\langle S_i^x S_j^x + S_i^y S_j^y \rangle}) - h \sum_{i\sigma} \sigma \lan \hat{n}_{i \sigma} \ran, \label{eq:W}
\eqx
%
but this is a nontrivial task, \tr{as after applying the Wick theorem too many terms appear (see Ref.~\onlinecite{Fukushima}, e.g. Eq.~(8) and the discussion afterwards), and one has to resort to making approximations at this point.} There are a few ways to perform this operation, and this is still an active field of research, so one can expect new calculation schemes to appear \cite{Bunemann}. Here, we use the scheme proposed recently by Fukushima \cite{Fukushima,Fukushima2} in the local-constraint version, which assumes that the average number of particles at any site and with any spin is unchanged by the projection,
\eq
\langle \hat{n}_{i\sigma} \rangle = \langle \hat{n}_{i\sigma} \rangle_0. \label{eq:local}
\eqx
This formalism is known to reproduce the Variational Monte Carlo results better than the conventional Gutzwiller approximation (at least, for the projected uniform nonmagnetic $d$-wave BCS superconductor - see Figs.   3 and 4 of Ref.~\onlinecite{Fukushima}). The local-constraint version of the formalism is quite general in the sense that it is capable of accounting for antiferromagnetism, superconductivity, and the ferromagnetic polarization. The explicit expressions for all averages appearing in Eq. (\ref{eq:W}) are given in Ref.~\onlinecite{Fukushima}. To express them in terms of mean-fields of our interest, we need to assume what is the character of the uncorrelated wave function $| \Psi_0 \ran$. Since our goal is the description of coexistence of AF and SC, we assume the corresponding mean-fields as nonzero at the level of $|\Psi_0 \ran$ as in the following. We start with the particle number in the form
\eq
n_{i\sigma} \equiv \lan \hat{n}_{i\s} \ran_0 = \half \Big( n + \sigma \, m_{FM} + \sigma \, m_{AF} \, e^{i \mathbf{Q r_i}} \Big),
\eqx
where $n$ is the band filling (assumed as constant), $m_{FM}$ is the ferromagnetic (longitudinal) spin-polarization component, and $m_{AF}$ is the antiferromagnetic (staggered) spin polarization. The factor $e^{i \mathbf{Q r_i}}$ (with $\bQ = (\pi, \pi)$) is responsible for the sign reversal of the staggered magnetic moment when exchanging the two sublattices A and B. \cite{Fazekas} We also assume the superconducting order parameter can be decomposed into two components
\eq
\Delta_{ij} \equiv \langle c_{j\da} c_{i\ua} \rangle_0 = \left\{
\begin{array}{ll} \tau_{ij} \Delta_A, &\textrm{for } i \in A-sublattice, \\
\tau_{ij} \Delta_B, &\textrm{for } i \in B-sublattice,
\end{array} \right.
\eqx
where $\tau_{ij}$ ensures the $d$-wave gap symmetry by setting $\tau_{ij} = +1 (-1)$ for $j = i \pm \hat{x}$ ($j = i \pm \hat{y}$) respectively, and with $\hat{x}$, $\hat{y}$ being the square-lattice basis vectors. The $d$-wave solution \tr{(of the $d_{x^2-y^2}$ form)} is taken throughout in the following analysis, \tr{as it is the most favorable energetically (cf. e.g. Ref.~\onlinecite{Ogata_tJ_Review})} The superconducting order parameter can be rewritten in terms of the singlet and the staggered $\pi$-triplet components, namely
\eqn
\Delta_{ij} &\equiv& \langle c_{j\da} c_{i\ua} \rangle_0 \equiv \half \Big( \langle c_{j\da} c_{i\ua} + c_{j\ua} c_{i\da} \rangle_0 + \langle c_{j\da} c_{i\ua} - c_{j\ua} c_{i\da}  \rangle_0 \Big) = \nonumber \\
 &=& \half \Big( \langle c_{j\da} c_{i\ua} - c_{i\da} c_{j\ua} \rangle_0 + \langle c_{j\da} c_{i\ua} + c_{i\da} c_{j\ua}  \rangle_0 \Big) \equiv \Delta^{(S)}_{ij} + \Delta^{(T)}_{ij} \, e^{i \mathbf{\bQ r_i}}, \label{eq:delta}
\eqnx
with
\eqn
\Delta^{(S)}_{ij} &\equiv& \half \tau_{ij} (\Delta_{A} + \Delta_{B}), \\
\Delta^{(T)}_{ij} &\equiv& \half \tau_{ij} (\Delta_{A} - \Delta_{B}).
\eqnx
The superconducting order parameter $\Delta_{ij}$ is defined on bond $\lan ij \ran$ (nearest-neighbor pair of sites). To define the gap per site, we make use of the standard \cite{Kyung} relation for $d$-wave solution
\eqn
\Delta_{i}^{(S)} &\equiv& \frac{1}{4} \sum_{j(i)} \tau_{ij} (\Delta_{i, j(i)} - \Delta_{j(i), i}) = \half (\Delta_{A} + \Delta_{B}), \\
\Delta_{i}^{(T)} &\equiv& \frac{1}{4} \sum_{j(i)} \tau_{ij} (\Delta_{i, j(i)} + \Delta_{j(i), i}) = \half (\Delta_{A} - \Delta_{B}) \, e^{i \mathbf{\bQ r_i}},
\eqnx
where $j(i)$ denotes the nearest neighbors of site $i$. The existence of the triplet component is inevitable even if there is no triplet channel in the pairing potential. Namely, the triplet component is dynamically induced by the singlet gap and antiferromagnetism. \cite{Psaltakis,Kyung,Tsonis,Aperis} From a microscopic point of view, this is also not surprising at all (see Fig.~\ref{fig1} for an intuitive illustration).
An interesting feature of the superconducting gap defined by Eq. (\ref{eq:delta}) is the nonzero momentum of Cooper pairs for the triplet component (it results from the $e^{i \mathbf{\bQ r_i}}$ term, in an analogy to center-of-mass momentum $\bQ$ in the Fulde-Ferrell-Larkin-Ovchinnikov (FFLO) phase \cite{FF,LO,*LO2, Matsuda-Rev}). A superconducting state with nonzero momentum has been investigated in a number of cases, \cite{Aperis, FF, LO,*LO2, Ogata_Q} even in zero external magnetic field. \cite{Loder, Maska2} The one presented here is analogous to that from Ref.~\onlinecite{Aperis}, except we consider both the microscopic $t$-$J$ model and the limit of strong correlations.

With the above assumptions, we can express the ground-state energy $W$ [cf. Eq. (\ref{eq:W})] as a function of the band filling $n$, the magnetization components $m_{FM}$ and $m_{AF}$, the superconducting gaps $\Delta_A$ and $\Delta_B$, and the hopping amplitudes $\chi_{ij\sigma}  \equiv \lan c_{i\sigma}\dg c_{j \sigma} \ran_0$. We assume as nonzero first- and second-nearest-neighbor hopping integrals $t$ and $t'$, which yields 6 different hopping amplitudes
\eq
\chi_{ij\sigma} \equiv \lan c_{i\sigma}\dg c_{j \sigma} \ran_0 \nolinebreak \in \nolinebreak \{\chi_{AB\ua}, \chi_{AB\da}, \chi_{AA\ua}, \chi_{AA\da}, \chi_{BB\ua}, \chi_{BB\da}\}.
\eqx
\tr{The resulting expression for $W$ is quite lengthy and has been presented in Appendix \ref{appA}.}
Next, as in the method proposed earlier \cite{JJ2,JJ1,Acta,SGA} (our present formulation is analogous to that from Ref.~ \onlinecite{JJ2}), to solve the model in a statistically-consistent manner, we impose additionally the constraints on all introduced mean fields by means of the Lagrange multipliers method. In effect, to carry out the subsequent analysis, we use the following energy operator
\eqn
\hat{K} &\equiv &
- \sum_{ij\s} [ \lambda_{ij\s}^{(\chi)} (c_{i\s}\dg c_{j\s} - \chi_{ij\s}) + h.c. ]
- \sum_{ ij } [ \lambda_{ij}^{(\Delta)} ( c_{j\da} c_{i\ua} - \Delta_{ij}) + h.c. ] \nonumber \\
&& - \sum_{ij\s} \lambda_{i\s}^{(n)} (\hat{n}_{i\s} - n_{i\s}) + W(n, m_{AF}, m_{FM}, \Delta_A, \Delta_B, \chi_{ij\s}) - \mu \sum_{i\s} \hat{n}_{i\s}. \label{eq:K}
\eqnx
This method of approach is equivalent in the $T \to \infty$ ($\beta \to 0$) limit to that presented in Refs.~\onlinecite{Wang} and \onlinecite{Yang}. The equivalence can be seen from the comparison of Eq. (\ref{eq:K}) and Eqs. (\ref{eq:l_first})-(\ref{eq:l_last}) with the corresponding equations from Refs.~\onlinecite{Wang} and \onlinecite{Yang} (e.g. Eq. (13) from Ref.~\onlinecite{Wang} provides an effective Hamiltonian with the operator part equivalent to our $\hat{K}$). The Lagrange multipliers $\lambda_{ij\s}^{(\chi)}$, $\lambda_{ij}^{(\Delta)}$, and $\lambda_{i\s}^{(n)}$ have the same symmetries, as the corresponding mean fields $\chi_{ij\s}$, $\Delta_{ij}$, and $n_{i\s}$. We also assume they are spatially homogeneous. Namely,
\eqn
\lambda_{i\s}^{(n)}     &\equiv& \lambda_n + \s \lambda_{m_{FM}} + \s \lm_{m_{AF}} \, e^{i \mathbf{\bQ r_i}} \\
\lambda_{ij}^{(\Delta)} &\equiv& \lambda_{\D}^{(S)} + \lambda_{\D}^{(T)} e^{i \mathbf{\bQ r_i}},
\eqnx
with
\eqn
\lambda_{\D}^{(S)} &=& \half(\lambda_{\D_A} + \lambda_{\D_B}), \\
\lambda_{\D}^{(T)} &=& \half(\lambda_{\D_A} - \lambda_{\D_B}).
\eqnx
After performing Fourier transformation of the operator part of $\hat{K}$ we obtain
\eqn
\hat{K} &=& \sum{}^{'}_{\bk} \Psi_{\bk}\dg \mb{M}_{\bk} \Psi_{\bk} + \Lambda (\mu + \lambda_n - \lambda_{m_{FM}}) + W(\vec{A}) \nonumber \\
&& + \Lambda \Big[n \lambda_n + m_{FM} \lambda_{m_{FM}} + m_{AF} \lambda_{m_{AF}} + 4 ( \Delta_A \lambda_{\Delta_A} + \Delta_B \lambda_{\Delta_B} ) \nonumber \\
&& + 4 \sum_\s ( 2 \chi_{AB\s} \lm_{\chi_{AB\s}} + \chi_{AA\s} \lm_{\chi_{AA\s}} + \chi_{BB\s} \lm_{\chi_{BB\s}}  )\Big],
\eqnx
where the primed summation runs over the folded (magnetic) Brillouin zone, by $\vec{A}$ we denote all the mean-fields, $\Lambda$ is the total number of sites, and the four-component operator $\Psi_{\bk}\dg$ has the following components
\eq
\Psi_{\bk}\dg = (c_{\bk\ua}\dg, c_{-\bk\da}, c_{\bk+\bQ\ua}\dg, c_{-\bk+\bQ\da}).
\eqx
The matrix $\mb{M}_{\bk}$ is given as
\eq
\mb{M}_{\bk} =\left(
\begin{array}{cccc}
  \xi_{\bk \ua} &
  -2 \lambda_{\D}^{(S)} \eta_{\bk}  &
   \zeta_{\bk+\bQ \ua} &      
    - 2 \lambda_{\D}^{(T)} \eta_{\bk+\bQ}
    \\
 -2 \lambda_{\D}^{(S)} \eta_{\bk}  &
  -\xi_{-\bk \da} &
   -2 \lambda_{\D}^{(T)} \eta_{\bk} &
     \zeta_{-\bk \da}      
    \\
  \zeta_{\bk \ua}  &      
  -2 \lambda_{\D}^{(T)} \eta_{\bk} &
    \xi_{\bk+\bQ \ua} &
    2 \lambda_{\D}^{(S)} \eta_{\bk+\bQ}
    \\
 -2 \lambda_{\D}^{(T)} \eta_{\bk+\bQ}  &
  \zeta_{-\bk+\bQ \da} &      
   2 \lambda_{\D}^{(S)} \eta_{\bk+\bQ}  &
    -\xi_{-\bk+\bQ \da}
\end{array}
\right), \label{eq:Mk}
\eqx
where
\eqn
 \zeta_{\bk \s} &=& - \s \, (\lm_{\chi_{AA\s}} - \lm_{\chi_{BB\s}}) \epsilon'_{\bk}-\lambda_{m_{AF}}, \\
 \xi_{\bk \s} &=& -\mu -\lm_n - \s \lm_{m_{FM}} - 2 \epsilon_{\bk} \lm_{\chi_{AB\s}} -  \epsilon'_{\bk} (\lm_{\chi_{AA\s}} + \lm_{\chi_{BB\s}}), \label{eq:dispersion} \\
  \epsilon_{\bk} &=& 2 (\cos{k_x} + \cos{k_y}), \\
  \epsilon'_{\bk} &=& 4 \cos{k_x} \cos{k_y}, \\
  \eta_\bk &=& \cos{k_x} - \cos{k_y}.
\eqnx
%
%
We have also used the fact that $\sum_\bk \epsilon_{\bk} = \sum_\bk \epsilon'_{\bk} = 0$. Note that in the present formulation $\lambda_{m_{FM}}$ corresponds to sum of the magnetic field $h$ and the correlation-induced field $h_{cor}$ \cite{JKJS,JKJS2} (or equivalently the Lagrange multiplier $\beta$ in the slave-boson theory \cite{Korbel, KR, SBREVIEW}). Namely, $\lambda_{m_{FM}} \equiv h+h_{cor} \equiv h+\beta$, what is evident from comparison of Eq. (\ref{eq:dispersion}) with appropriate expressions taken from Refs.~\onlinecite{JKJS, JKJS2, Korbel}.

Next, we determine the eigenvalues of $\mb{M}_{\bk}$, as they correspond to quasiparticle excitations of the system. An analytic diagonalization of $\mb{M}_{\bk}$ produces very long expressions, and more importantly, expressions with square roots of possibly negative numbers. Therefore, having in mind their subsequent implementation to calculate the physical properties, we diagonalize this matrix numerically. Next, having determined the eigenvalues $\{E_{\bk i}\}_{i=1, 2, 3, 4}$, we determine the generalized grand potential functional for the system of fermions, which is
\eqn
\mathcal{F} &=& -\beta^{-1} \sum{}^{'}_{\bk, i=1, 2, 3, 4} \ln(1 + e^{-\beta E_{\bk i}}) + \Lambda (\mu + \lambda_n - \lambda_{m_{FM}}) + W(\vec{A}) \nonumber \\
&& + \Lambda \Big[n \lambda_n + m_{FM} \lambda_{m_{FM}} + m_{AF} \lambda_{m_{AF}} + 4 ( \Delta_A \lambda_{\Delta_A} + \Delta_B \lambda_{\Delta_B} ) \nonumber \\
&& + 4 \sum_\s ( 2 \chi_{AB\s} \lm_{\chi_{AB\s}} + \chi_{AA\s} \lm_{\chi_{AA\s}} + \chi_{BB\s} \lm_{\chi_{BB\s}}  )\Big].
\eqnx
The physical (equilibrium) values of the mean fields and the Lagrange multipliers are obtained from the necessary conditions for $\mathcal{F}$ to have a minimum subject to the constraints, i.e.,
\begin{equation}
\frac{\partial \mathcal{F}}{\partial \vec{A}} = 0, ~~~~~ \frac{\partial \mathcal{F}}{\partial \vec{\lambda}} = 0,
\label{derivative of mathcalF A, lambda}
\end{equation}
where by $\vec{\lm}$ we denote collectively the Lagrange multipliers. Equations $\partial \mathcal{F} / \partial \vec{A} = 0$ provide the explicit analytic expressions for the Lagrange multipliers, i.e.,
\eqn
\lm_{n} &=& -\Lambda^{-1} \partial_{n} W(\vec{A}), \label{eq:l_first} \\
\lm_{m_{FM}} &=& -\Lambda^{-1} \partial_{m_{FM}} W(\vec{A}), \\
\lm_{m_{AF}} &=& -\Lambda^{-1} \partial_{m_{AF}} W(\vec{A}), \\
\lm_{\D_{A}} &=& -\frac{1}{4} \Lambda^{-1} \partial_{\D_{A}} W(\vec{A}), \\
\lm_{\D_{B}} &=& -\frac{1}{4} \Lambda^{-1} \partial_{\D_{B}} W(\vec{A}), \\
\lm_{\chi_{AB\s}} &=& -\frac{1}{8} \Lambda^{-1} \partial_{\chi_{AB\s}} W(\vec{A}), \\
\lm_{\chi_{AA\s}} &=& -\frac{1}{4} \Lambda^{-1} \partial_{\chi_{AA\s}} W(\vec{A}), \\
\lm_{\chi_{BB\s}} &=& -\frac{1}{4} \Lambda^{-1} \partial_{\chi_{BB\s}} W(\vec{A}). \label{eq:l_last}
\eqnx
The above expressions can be utilized to eliminate Lagrange multipliers $\vec{\lm}$ from the solution procedure. Thus, we obtain 11 equations to be solved numerically for the mean fields $\vec{A}$, instead of 22 equations for both $\vec{A}$ and $\vec{\lm}$. The equations for the mean fields (obtained from $\partial \mathcal{F} / \partial \vec{\lambda} = 0$) have the following form
\eqn
0 &=& \beta^{-1} \partial_{\lm_n} f_\beta(\vec{\lm})  - \Lambda(n-1), \label{eq:s_first} \\
0 &=& \beta^{-1} \partial_{\lm_{m_{FM}}} f_\beta(\vec{\lm})  - \Lambda(m_{FM}+1), \\
0 &=& \beta^{-1} \partial_{\lm_{m_{AF}}} f_\beta(\vec{\lm})  - \Lambda m_{AM} , \\
0 &=& \beta^{-1} \partial_{\lm_{\D_A}} f_\beta(\vec{\lm})  - 4 \Lambda \D_A , \\
0 &=& \beta^{-1} \partial_{\lm_{\D_B}} f_\beta(\vec{\lm})  - 4 \Lambda \D_B , \\
0 &=& \beta^{-1} \partial_{\lm_{\chi_{AB\s}}} f_\beta(\vec{\lm}) - 8 \Lambda \chi_{AB\s} , \\
0 &=& \beta^{-1} \partial_{\lm_{\chi_{AA\s}}} f_\beta(\vec{\lm}) - 4 \Lambda \chi_{AA\s} , \\
0 &=& \beta^{-1} \partial_{\lm_{\chi_{BB\s}}} f_\beta(\vec{\lm}) - 4 \Lambda \chi_{BB\s}, \label{eq:s_last}
\eqnx
where
\eq
f_\beta(\vec{\lm}) \equiv \sum_{\bk, i=1..4} \ln(1 + e^{-\beta E_{\bk i}}). \label{eq:f_beta}
\eqx
The derivative $\partial_{\lm_n} f_\beta(\vec{\lm})$ is computed numerically with a 5-point stencil method (as it gives two-three orders of magnitude better precision than the standard 3-point stencil). For example
\eqn
\partial_{\lm_n} f_\beta(\vec{\lm}) &=& \frac{1}{12 x} \Big[ -f_\beta(\lm_n + 2 x, \lm_{mFM}, \lm_{mAF}, ...) + 8 f_\beta(\lm_n + x, \lm_{mFM}, \lm_{mAF}, ...) \nonumber \\
&& - 8 f_\beta(\lm_n - x, \lm_{mFM}, \lm_{mAF}, ...) + f_\beta(\lm_n - 2 x, \lm_{mFM}, \lm_{mAF}, ...)\Big] \nonumber \\
&& + O(x^4) \label{eq:derivative}
\eqnx
where we use the ``equilibrium'' values of $\vec{\lm}$ as given by Eqs. (\ref{eq:l_first})-(\ref{eq:l_last}). The step $x$ is typically equal to $x=0.0001$. Larger values of $x$ would cause greater error in the above formula. Smaller values would result in a loss of numerical precision. \tr{We have verified that at $h=0$ (where analytical formulas for the eigenvalues $E_{\bk i}$ are available) the numerical computation of the derivatives according to Eq. (\ref{eq:derivative}) with the chosen step $x=0.0001$ introduces error smaller than the precision of the procedure of solving the set of Eqs. (\ref{eq:s_first})-(\ref{eq:s_last}). The value of the step $x$ has been chosen after an analysis of the error at $h=0$ and the numerical-precision loss (for very small $x$ the numerical-precision loss lead to impossibility of solving the set of Eqs. (\ref{eq:s_first})-(\ref{eq:s_last}) with the given precision).}

\section{Results and physical discussion} \label{sec:results}

The equations (\ref{eq:s_first})-(\ref{eq:s_last}) are solved numerically with the use of GNU Scientific Library (GSL) \cite{GSL} on a grid of size $\Lambda = 256\, \times \,256$. \tr{We use the} \verb"gsl_multiroot_fsolver_hybrids" \tr{solver which implements the hybrids algorithm. We use the precision $epsabs = 10^{-7}$. Namely, the procedure converges when the relation $\sum_i |f_i| < epsabs$ is fulfilled (where the sum is taken over all equations, which have been brought to the form $f_i = 0$ and divided by $\Lambda$ to ensure lattice-independent convergence conditions).} We assume the following values of parameters: $t=3$, $t'=t/4=0.75$, $J=1$, and $\beta=500$, what yields the temperature $T=1/\beta=0.002 \approx 0$. In Table I the exemplary numerical values of the parameters have been provided for the sake of completeness. Numerical accuracy is on the level of the last digit specified. The energy scale has been set by taking the value of the exchange integral as unit, $J=1$. \tr{For more details on the numerical procedure see Chapter~6 of Ref.~\onlinecite{JKPHD}.}

\begin{center}
\begin{tabular}{| c | c || c | c |}
\hline
  \multicolumn{4}{|c|}{Table I. Equilibrium values of mean-field variables, Lagrange multipliers, } \\
  \multicolumn{4}{|c|}{free energy $F$ and grand potential functional $\mathcal{F}$ at $h = 0.3$ and $\beta=500$.} \\
\hline
Variable & Value & Variable & Value  \\ \hline
$ \mu $          & 3.2997360 	   & $ \lm_n $                & -5.1331996      \\
$ m_{FM} $ 		 & 0.0000000       & $ \lm_{m_{FM}} $ 		& 0.3000001      \\
$ m_{AF} $       & 0.8100315       & $ \lm_{m_{AF}} $         & 2.5817963       \\
$ \Delta_A $     & 0.0922998       & $ \lm_{\Delta_A} $       & 0.2730140      \\
$ \Delta_B $     & 0.0479298       & $ \lm_{\Delta_B} $       & 0.4395630       \\
$ \chi_{AB\ua} $ & 0.1218625       & $ \lm_{\chi_{AB\ua}} $   & 0.4258402      \\
$ \chi_{AB\da} $ & 0.1218625       & $ \lm_{\chi_{AB\da}} $   & 0.4258402      \\
$ \chi_{AA\ua} $ & -0.0167505      & $ \lm_{\chi_{AA\ua}} $   & -0.1031297     \\
$ \chi_{AA\da} $ & 0.0275895       & $ \lm_{\chi_{AA\da}} $   & -0.0120561    \\
$ \chi_{BB\ua} $ & 0.0275895       & $ \lm_{\chi_{BB\ua}} $   & -0.0120561    \\
$ \chi_{BB\da} $ & -0.0167505      & $ \lm_{\chi_{BB\da}} $   & -0.1031297     \\
$ F /\Lambda$    & -1.0110048      & $ \mathcal{F}/\Lambda $  & -4.2117488      \\ \hline \hline
\end{tabular}
\end{center}

A number of stable phases emerge as solutions of the equations, depending on the physical condition ($n$, $h$). As we work with constant number of particles $n$, the stable phase is the one with the lowest free energy, defined by
\eq
F = \mathcal{F}_0 + \mu n \Lambda, \label{eq:FE}
\eqx
where all the optimal values of mean fields and Lagrange multipliers (i.e. those being solution to Eqs. (\ref{eq:l_first})-(\ref{eq:s_last})) are inserted in the functional $\mathcal{F}$ and $\mu$ is the chemical potential.
\begin{figure}
\begin{center}
\includegraphics[width=9cm,angle=270]{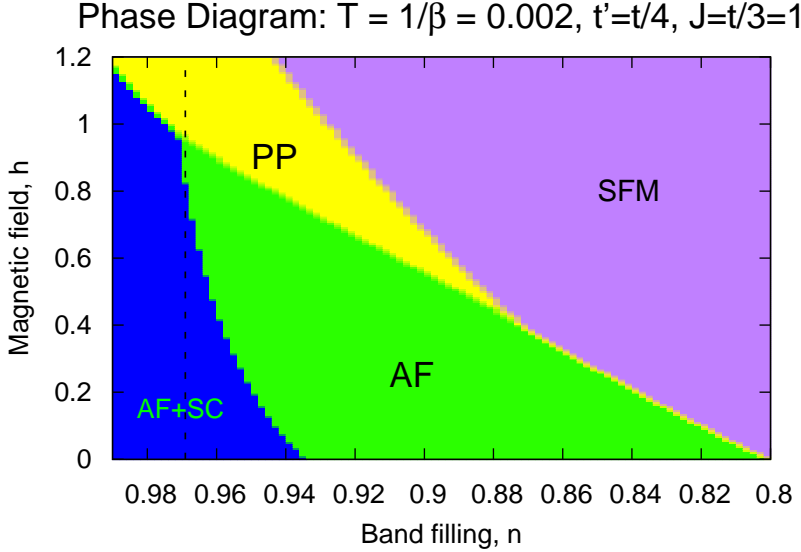}
\end{center}
\caption{(Color online)  Phase diagram on the band filling - magnetic field plane. The phases are labeled as follows: AF+SC - phase with coexisting superconductivity and antiferromagnetism, AF - antiferromagnetic phase, \tr{PP - polarized paramagnetic phase}, SFM - saturated ferromagnetic phase (with $m_{FM} = n$). No stable pure superconducting solution has been found. For a further analysis we restrict ourselves to $n=0.97$ as marked by the dashed vertical line.}
\label{fig2}
\end{figure}

The exemplary phase diagram on the band filling $n$ - magnetic field $h$ plane is exhibited in Fig.~\ref{fig2}. It can be seen that antiferromagnetic phase is the predominant one in the low-field regime and above $n=0.8$. For $n \gtrsim 0.935$ antiferromagnetism coexists with superconductivity, what amounts to a phase with nonzero three order parameters (similarly, as in e.g. Ref.~\onlinecite{Aperis}). In the low-$n$ part of the phase diagram (for $n<0.8$) the saturated ferromagnetic (SFM) phase with $m_{FM} = n$ becomes the stable state. This phase is stable even in the $h \ra 0$ limit. This is an interesting result, which adds to the discussion of ferromagnetism in the $t$-$J$ \cite{Koretsune,*Koretsune2,Putikka} model. 
There is also a number of papers (see e.g. Refs.~\onlinecite{JJ2} and \onlinecite{Marcin1}) analyzing the $t$-$J$ model (\ref{eq:tJ}) with the Gutzwiller-type of approach with the parameters in a similar range (i.e. with $n~<~0.8$ and similar values of $t_{ij}$ and $J$). Some of those papers disregard completely the Zeeman-term influence, and this omission is justified when applying the model to high-Tc superconductors, where orbital effects dominate over the Pauli magnetism. We have shown, that even at $h=0$ the system may be completely spin-polarized and therefore, the inclusion of ferromagnetic polarization $m_{FM}$ is important in treatment of the $t$-$J$ model. \tr{Finally, our phase diagram can be compared (although this is not a direct comparison, as even at $h=0$ we have $m_{FM} \neq 0$ for the AF phase) to that obtained by other Gutzwiller approximation scheme, \cite{Ogata} in which the coexisting phase was stable up to the doping $\delta \equiv 1-n = 0.1$, and for higher doping levels the pure superconducting state was stable. In our approach the antiferromagnetic phase is stable for such dopings instead. We comment on the strong antiferromagnetism in the following analysis.}

At band filling $n=0.97$ the phase diagram (or the phase sequence as a function of field $h$) resemble those observed recently in the heavy-fermion compounds CeCo(In$_{1-x}$Cd$_{x}$)$_5$ \cite{Nair} at doping $x=0.0075$ and CeRhSi$_3$ \cite{Kimura} at pressure $p~\approx~17\textrm{ kbar}$.\footnote{Although in CeRhSi$_3$ the phases have only been analyzed as a function of temperature.} Namely, in low magnetic fields a phase with coexisting antiferromagnetic and superconducting orders (AF+SC) is stable, whereas for higher magnetic fields a continuous transition to a pure antiferromagnetic (AF) phase takes place, followed by a discontinuous transition to the \tr{polarized paramagnetic (PP)} phase. The phases appearing at this band filling ($n=0.97$) are analyzed thus in detail in the following.

\begin{figure}
\begin{center}
\includegraphics[width=9cm,angle=270]{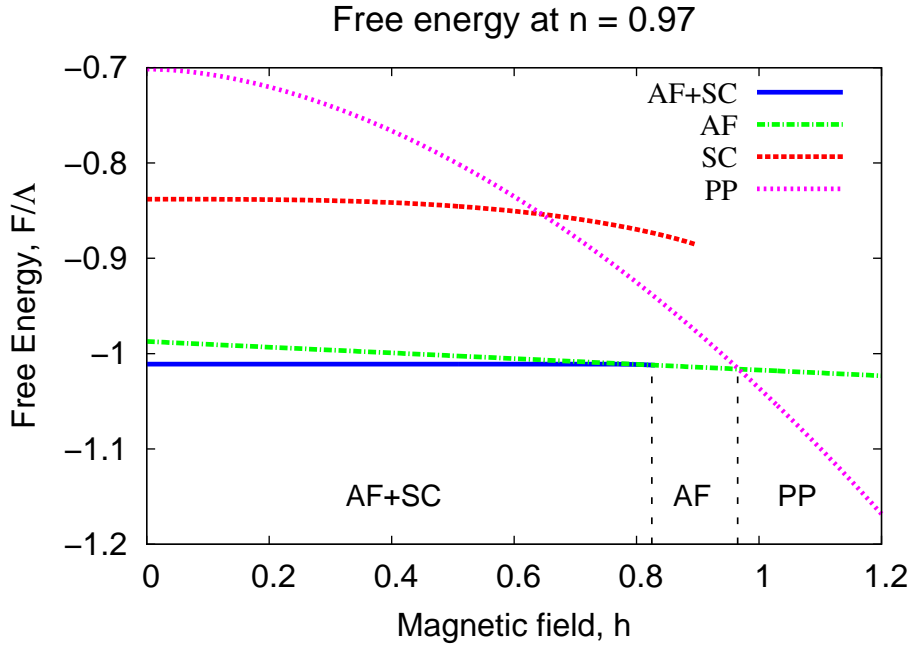}
\end{center}
\caption{(Color online) Free energy per site as a function of magnetic field for the specified selection of phases. Types of order are marked explicitly at the bottom. The ``SC'' phase is the pure superconducting phase (i.e. with $m_{AF}=0$), which obviously has higher energy than other phases and hence does not appear in the phase diagram. The vertical dashed lines mark the phase boundaries: the AF+SC - AF line marks a continuous transition, whereas that for the AF - PP is discontinuous, as one can see by looking at the behavior of the slope $\partial F/\partial h$.}
\label{fig3}
\end{figure}

In Fig.~\ref{fig3} we have plotted the free energy curves for a choice of possible {\it a priori} phases. It can be seen (also from the following Figures) that the transition AF+SC $\ra$ AF is continuous, whereas the transition AF $\ra$ PP is of the first order. Also, pure superconducting (SC) solution is unstable, and this holds for other band fillings as well. It can be concluded from Fig.~\ref{fig3} that antiferromagnetism is the ``dominating'' phenomenon, since the energy gain from developing antiferromagnetic order (which can be seen from closer look at the difference ($F_{PP} - F_{AF}$)) is much higher than the gain from developing superconducting order ($F_{PP} - F_{SC}$). Moreover, the energy gain from developing AF order within SC phase ($F_{SC} - F_{AF+SC}$) is much higher than that from developing SC order within the AF phase ($F_{AF} - F_{AF+SC}$). \tr{This observation can be compared to the results of the Variational Monte Carlo method, \cite{Himeda} in which the $d$-wave solution is only slightly higher in energy than the coexisting phase (see Fig. 2 of Ref.~\onlinecite{Himeda}).}

\begin{figure}
\begin{center}
\includegraphics[width=9cm]{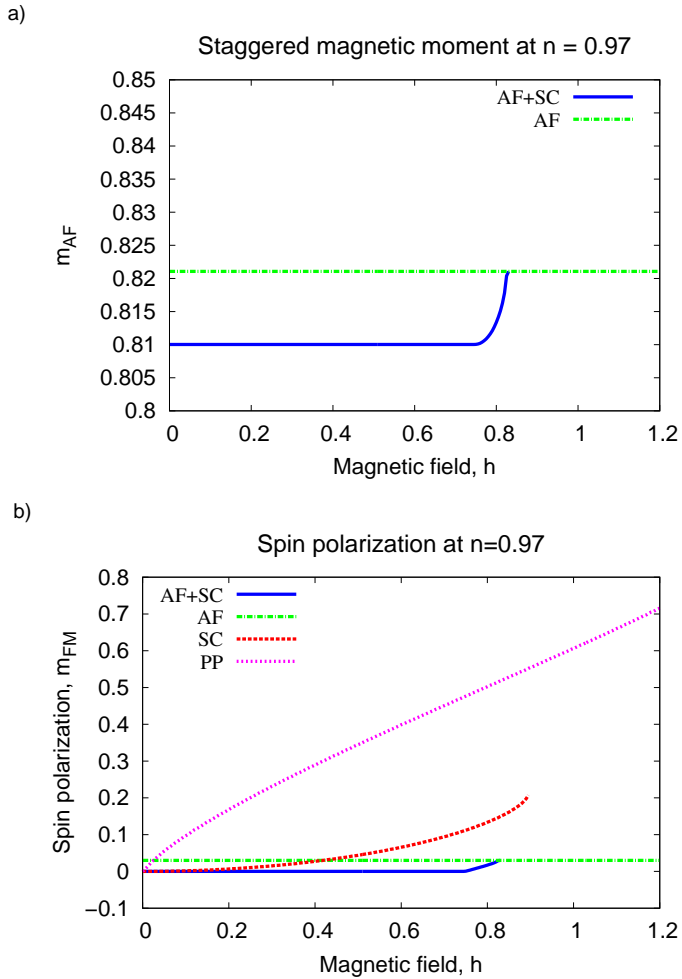}
\end{center}
\caption{(Color online) Staggered magnetic moment (a) and ferromagnetic spin polarization (b) for the selected phases. Obviously, the staggered moment of the SC and PP phases is 0, and has not been plotted in (a). The magnetic moment value is insensitive to the projection (i.e., it is the same in both the correlated $|\Psi\ran$ and the uncorrelated $|\Psi_0\ran$ states).}
\label{fig4}
\end{figure}

In Fig.~\ref{fig4} we exhibit the magnetic moment per site of the system for different phases. Namely, we plot the staggered magnetization $m_{AF}$ and the ferromagnetic magnetization $m_{FM}$ (spin-polarization). The staggered magnetization is close to the saturation value of $m_{AF} = n = 0.97$. \tr{Such overestimation of the staggered magnetization value over the Variational Monte Carlo results \cite{Himeda, Giamarchi} is also present in the slave-boson approach. \cite{Inaba} This is not surprising, as the method we use \cite{JJ2,JJ1,Acta,SGA} is similar in structure (the Lagrange multipliers are present in both methods) to the slave-bosons approach (for the discussion of the equivalence for the paramagnetic state see Ref.~\onlinecite{SGA}).} The obtained ferromagnetic spin-polarization for the pure AF phase is equal to $m_{FM} = 1-n$ at all magnetic fields. Also, it can be seen that development of the SC order within the AF phase alters by a small amount the staggered magnetization $m_{AF}$, which drops by approximately $1\%$ (see Fig.~ \ref{fig4}a).

\begin{figure}
\begin{center}
\includegraphics[width=7.3cm]{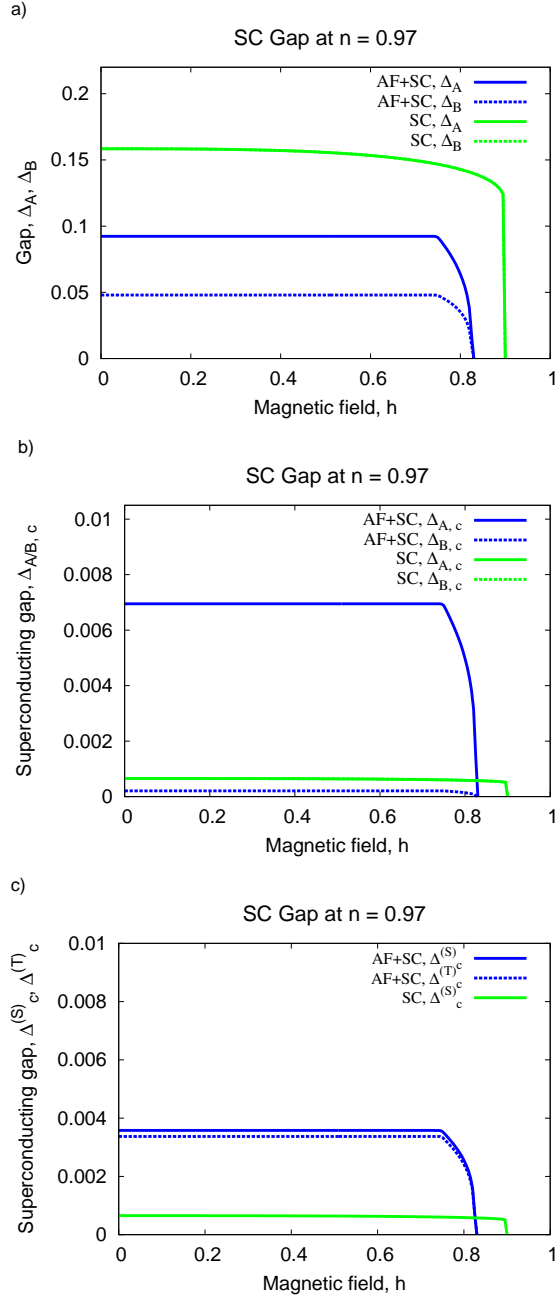}
\end{center}
\caption{(Color online) Superconducting gaps versus magnetic field for AF+SC and SC phases. (a) $\Delta_A$ and $\Delta_B$ gaps obtained for the uncorrelated wave function $|\Psi_0\ran$; (b) and (c) gaps for the state specified by the correlated wave function (labeled as $\Delta_{c}$), as given by Eq. (\ref{eq:ave}). (b) shows the sublattice-specific $\Delta_A$ and $\Delta_B$ gaps, and (c) shows the singlet and triplet components of the gap. Note that the superconducting gaps ($\Delta_{c}$) are enhanced in the AF+SC state (with respect to that in the pure SC state, which is however an unstable state). Also, the singlet and the staggered-triplet components are almost equal in the correlated state.}
\label{fig5}
\end{figure}

In Fig.~\ref{fig5} various superconducting gaps are shown. Namely we exhibit both the ``uncorrelated'' gap for the wave function $|\Psi_0 \ran$, as well as the gap for the correlated wave function $|\Psi \ran$, the latter defined by Eq. (\ref{eq:ave}) and labeled as $\Delta_{c}$. In the pure SC phase the sublattice gaps are equal ($\Delta_A = \Delta_B$), what amounts to the absence of the triplet component. Note that although the uncorrelated gaps ($\Delta_{A}$, $\Delta_{B}$) are larger in the pure SC phase than in the AF+SC phase, the correlated gaps ($\Delta^{(S)}_{c}$, $\Delta^{(T)}_{c}$) are much larger in the AF+SC phase than in the pure SC phase. This very important conclusion means that the presence of antiferromagnetism supports superconductivity in the present situation. The opposite is not true as the staggered moment is slightly larger in the AF phase than in the AF+SC phase. Finally, the renormalized gaps are more than an order of magnitude smaller than their bare (uncorrelated) correspondants.

The picture with large antiferromagnetic magnetization $m_{AF}$ (Fig.~\ref{fig4}) and small superconducting gap (Fig.~\ref{fig5}) is consistent with the energy curves displayed in Fig.~\ref{fig3}. To shift the energy balance towards the SC phase one could either decrease $t'$, or increase $J$. \tr{By doing that within a wide parameter margin, the antiferromagnetic phase still remains a predominant phase. Other possibility to weaken antiferromagnetism is to include additionally the intersite attraction ($V \sum_{\lan i j \ran} \hat{n}_i \hat{n}_j$) in the starting Hamiltonian. This has been shown to stabilize the $d$-wave superconducting state \cite{Yanase} (see Ref.~\onlinecite{Fukushima2} for the expression for the average value of this term within the extended Gutzwiller scheme we use). The strong antiferromagnetism may represent an apparent feature of the Gutzwiller scheme used}, \cite{Fukushima} in which magnetization is not changed by the projection, as follows directly from Eq.~(\ref{eq:local}).

\begin{figure}[h!]
\begin{center}
\includegraphics[width=15cm]{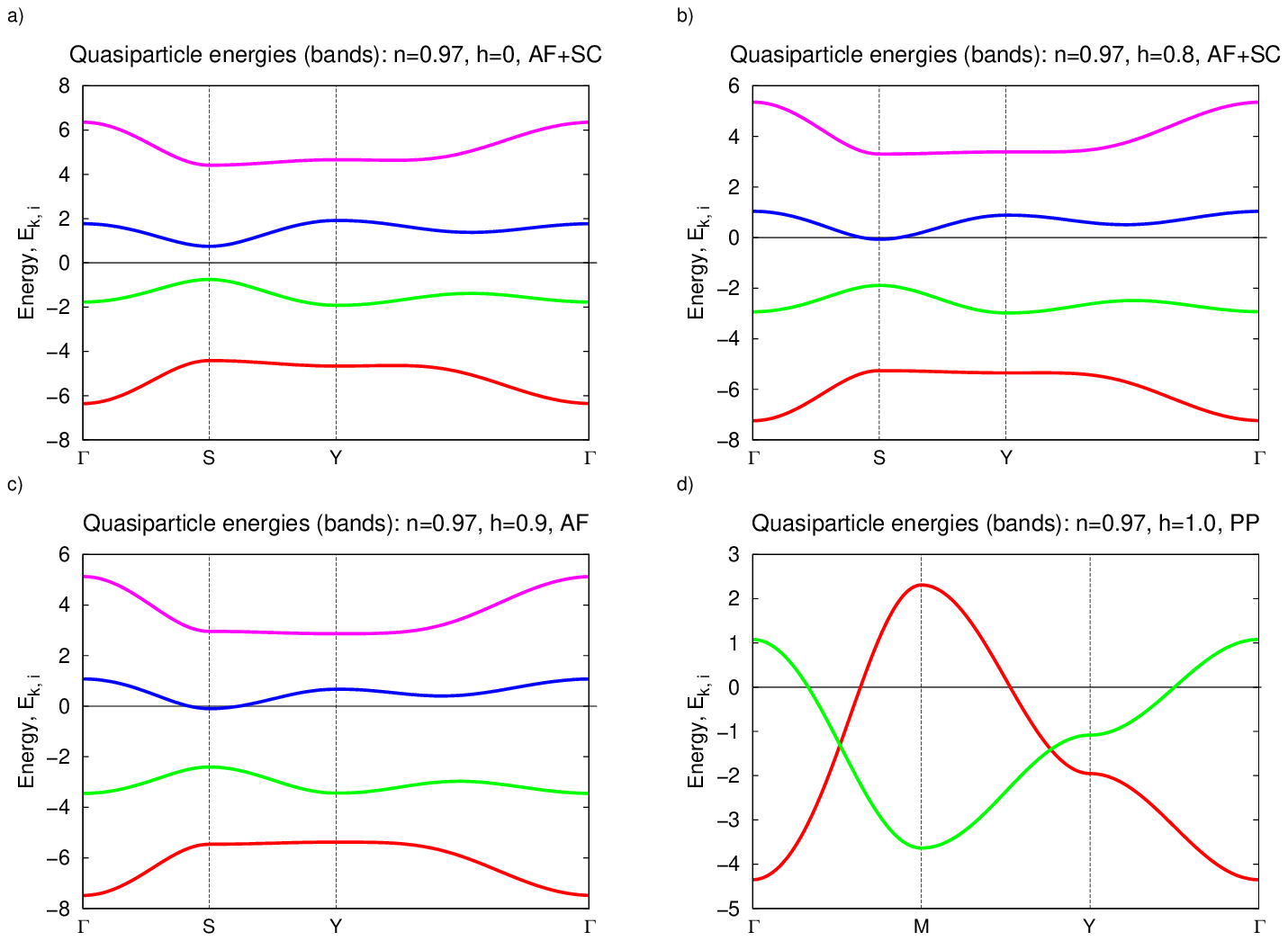}
\end{center}
\caption{(Color online) Quasiparticle energies (bands) for phases obtained at $n=0.97$: (a) AF+SC phase ($h=0$), (b) AF+SC phase with nonzero spin-polarization ($h>0$), (c) AF phase, and (d) PP phase (for a different path in the Brillouin zone). \tr{The full Brillouin zone (d) is spanned by the vertices $(\pm \pi, \pm \pi)$, whereas the folded (magnetic) Brillouin zone (a-c), by $(\pm \pi, 0)$ and $(0, \pm \pi)$. The characteristic points are the following: $\Gamma = (0, 0)$, $S = (\pi/2, \pi/2)$, $Y = (0, \pi)$, and $M = (\pi, \pi)$.} Note that in (a-c) two of the energies $E_{\bk i}$ (those with $E_{\bk i}>0)$ describe quasiparticle states, and the other two represent quasihole states. Also, in (d) only two energies are displayed, as the Brillouin zone is not the folded (magnetic) one, but the full Brillouin zone, because there is no antiferromagnetism in this case. The red (dark) curve in (d) describes a quasiparticle with $\s=\ua$, and the green curve describes a quasihole with $\s=\da$. The fully gapped electronic structure in (a)-(c) is caused by the magnetic (renormalized Slater) gap appearance in the AF+SC and the AF phases. Energy scale is in units of $J$.}
\label{fig6}
\end{figure}

Finally, in Fig.~\ref{fig6} we display the quasiparticle energies (the Slater subbands) for the phases discussed above for $n=0.97$. The crossing of one of the bands with the zero energy line at the S point of the Brillouin zone in Figs.~\ref{fig6}bc means that the quasiparticle excitations will be spontaneously created (are gapless), a circumstance leading to a nonzero spin-polarization (cf. Fig.~\ref{fig6}), similarly as in the situation for the FFLO state. \cite{FF} Note also that for $h=0$ the AF+SC electronic structure is gapful for the $d$-wave superconducting phase, because of the presence of the Slater-type (magnetic) splitting. This is not true anymore for $h \gtrsim 0.8$ (cf. Figs. \ref{fig6}bc and Fig. \ref{fig4}a) when a uniform ferromagnetic component appears. Also, the bands are sizably wider in the PP state.

\section{Summary} \label{sec:summary}

In summary, we have carried out a detailed analysis of the coexistence of antiferromagnetism and superconductivity within a microscopic $t$-$J$ model, with the Zeeman term included. The strong correlations were accounted for by means of the extended Gutzwiller projection method. Also, the constraints assuring the statistical-consistency of the results have been included. We have obtained the phase diagram on the band filling-magnetic field plane, in which for the band fillings in the range $n\approx 0.935 - 0.970$ and with the increasing magnetic field, a series of phase transitions takes place. Namely, the system evolves from the coexisting (AF+SC) phase, through the antiferromagnetic (AF) phase, to the \tr{polarized paramagnetic (PP) phase}. Also, the onset of superconducting order reduces the AF order parameter. By contrast, the superconducting gaps are enhanced by the presence of the AF order. In the AF+SC phase there are two superconducting gaps of an almost equal amplitude: the singlet and the staggered-triplet components. These features reflect in a qualitative manner the experimental findings in the CeCo(In$_{1-x}$Cd$_{x}$)$_5$ \cite{Nair, Urbano} and CeRhSi$_3$ \cite{Kimura} heavy fermion systems, \tr{although our model is too simplified to be quantitatively related to such complex heavy fermion systems.} Additionally, both antiferromagnetism and superconductivity originate from the same electrons and are driven by the same kinetic-exchange interaction.
Note that the real-space pairing is the pairing without ``boson glue'', i.e. without {\it paramagnons}. It is the mechanism of pairing arising entirely from interelectron correlations which is particularly effective when renormalized hopping and exchange interaction are of comparable magnitude.

As mentioned earlier, it would be very interesting to perform similar analysis within the Periodic Anderson Model, as this might allow for a comparison with the experiments [work along this line is in progress in our group \cite{Olga}]. Also, testing other Gutzwiller schemes seems crucial to verify if the strong antiferromagnetism and the tendency towards saturated ferromagnetism are only the characteristic features of the utilized renormalization scheme, or represent a universal tendency of the projected $t$-$J$ model. For that purpose, the inclusion of realistic, orbitally degenerate $f$ level structure, not just pseudospin $\Gamma_7$ doublet of Ce$^{3+}$, would be desirable.

One should also note that the present approach includes the effect of applied magnetic field only via Zeeman term (the Pauli limit). For discussion of high-temperature superconductivity for $h>0$ the orbital effects should be incorporated.

\section*{Acknowledgements}
The work was supported by Ministry of Higher Education and Science, Grants Nos. N N202 173735 and N N202 128736, as well as by the Foundation for Polish Science under the "TEAM" program for the years 2011-2014. Discussions with Olga Howczak and Jakub J\c{e}drak are appreciated.

\appendix

\section{Explicit expression for $W$} \label{appA}

We provide here the expression for $W \equiv \langle \hat{\mathcal{H}}_{tJ} \rangle_0$. This expression can be divided into parts coming from different terms of Hamiltonian with $W_t \equiv \sum_{ij\sigma} t_{ij} \langle c_{i\sigma}\dg c_{j\sigma} \rangle$ and $ W_J \equiv J \sum_{\langle ij \rangle} ({\langle S_i^z S_j^z \rangle} + {\langle S_i^x S_j^x + S_i^y S_j^y \rangle})$, as follows $W = W_t + W_J - \Lambda h m_{FM}$.
The expressions for $W_J$ and $W_t$ are given by

\eqn
W_J &=& 2 J \Lambda (-\frac{4 (\chi_{AB\da} \chi_{AB\ua} + \Delta_A \Delta_B)}{\sqrt{
    m_{AF}^4 + (m_{FM}^2 - (-2 + n)^2)^2 - 2 m_{AF}^2 (m_{FM}^2 + (-2 + n)^2)}} \nonumber \\
   && + \frac{1}{4} ((-m_{AF} + m_{FM}) (m_{AF} + m_{FM}) \nonumber \\
   && - (4 (\Delta_A^2 (-1 + m_{AF} - m_{FM}) (-1 + m_{AF} + m_{FM}) \nonumber \\
   && \times (2 + m_{AF} - m_{FM} - n) (2 + m_{AF} + m_{FM} - n) \nonumber \\
   && + \chi_{AB\ua}^2 (1 + m_{AF} - m_{FM}) (-1 + m_{AF} + m_{FM}) \nonumber \\
   && \times (2 + m_{AF} + m_{FM} - n) (-2 + m_{AF} - m_{FM} + n) \nonumber \\
   && + \chi_{AB\da}^2 (-1 + m_{AF} - m_{FM}) (1 + m_{AF} + m_{FM}) \nonumber \\
   && \times (2 + m_{AF} - m_{FM} - n) (-2 + m_{AF} + m_{FM} + n) \nonumber \\
   && + \Delta_B^2 (1 + m_{AF} - m_{FM}) (1 + m_{AF} + m_{FM}) \nonumber \\
   && \times (-2 + m_{AF} - m_{FM} + n) (-2 + m_{AF} + m_{FM} + n))) \nonumber \\
   && /((2 + m_{AF} - m_{FM} - n) (2 + m_{AF} + m_{FM} - n) \nonumber \\
   && \times (-2 + m_{AF} - m_{FM} + n) (-2 + m_{AF} + m_{FM} + n))))
\eqnx

and 

\eqn
W_t &=& 2 \Lambda (-4 (1 - n) (\frac{4 \chi_{AB\da} \Delta_A \Delta_B + \chi_{AB\ua} (4 \chi_{AB\da}^2 + m_{AF}^2 - (2 + m_{FM} - n)^2)}{(m_{AF}^2 - (2 + m_{FM} - n)^2) \sqrt{-m_{AF}^2 + (-2 + m_{FM} + n)^2}} \nonumber \\
   && + \frac{4 \chi_{AB\ua} \Delta_A \Delta_B + \chi_{AB\da} (4 \chi_{AB\ua}^2 + m_{AF}^2 - (-2 + m_{FM} + n)^2)}{\sqrt{-m_{AF}^2 + (2 + m_{FM} - n)^2} (m_{AF}^2 - (-2 + m_{FM} + n)^2)}) t \nonumber \\
   && + (2 (-1 + n) (-\frac{(\chi_{BB\ua} (-4 \chi_{BB\da}^2 + (-2 + m_{AF} - m_{FM} + n)^2))}{(2 + m_{AF} - m_{FM} - n)} \nonumber \\
   && + \frac{\chi_{BB\da} (-2 + m_{AF} - m_{FM} + n) (-4 \chi_{BB\ua}^2 + (-2 - m_{AF} + m_{FM} + n)^2)}{-2 - m_{AF} + m_{FM} + n}^2) t') \nonumber \\
   && /(-2 + m_{AF} - m_{FM} + n)^2 \nonumber \\
   && + (2 (-1 + n) (\chi_{AA\ua} (-4 \chi_{AA\da}^2 + (2 + m_{AF} + m_{FM} - n)^2) (-2 + m_{AF} + m_{FM} + n) \nonumber \\
   && - \chi_{AA\da} (2 + m_{AF} + m_{FM} - n) (-4 \chi_{AA\ua}^2 + (-2 + m_{AF} + m_{FM} + n)^2)) t')\nonumber \\
   && / ((2 + m_{AF} + m_{FM} - n)^2 (-2 + m_{AF} + m_{FM} + n)^2)).
\eqnx

\bibliography{AFFFLO}

\end{document}